\documentclass[anonymous=false, %
               format=acmsmall, %
               screen=true, %
               nonacm=true]{acmart}

\usepackage[ruled]{algorithm2e} 
\newcommand{\por}{\,\vert\,}
\newcommand{\Par}{\,\Vert\,}
\DeclareMathOperator*{\argmin}{argmin}

\usepackage{listings}
\usepackage{cleveref}
\usepackage{todonotes}
\usepackage{pifont}
\usepackage{adjustbox}
\usepackage{array}

\newcommand{\Green}[1]{\color{green!50!black}{#1}}
\newcommand{\Red}[1]{\color{red!60!black}{#1}}
\newcommand{\emark}{{\color{red!60!black}\ding{55}}}

\lstnewenvironment{lstpython}%
{\lstset{language=Python}}
{}
\lstnewenvironment{lstpythontable}%
{\lstset{language=Python,basicstyle=\small\ttfamily,morekeywords={as},numbers=left,xleftmargin=0em}}
{}
\lstnewenvironment{lstpythons}%
{\lstset{language=Python,basicstyle=\ttfamily\small}}
{}
\def\python{\lstinline[language=Python, basicstyle=\normalsize\ttfamily]}

\lstdefinelanguage{Stan}[]{}{
  morekeywords=[3]{guide,parameters,networks,variational},%
  morekeywords=[2]{functions,model,data,parameters,quantities,%
    transformed,generated},
  morekeywords=[1]{for,in,while,repeat,until,if,then,else,%
  true,false,%
  int,real,vector,simplex,ordered,positive_ordered,%
  row_vector,unit_vector,matrix,cholesky_factor_corr, cholesky_factor_cov,%
  corr_matrix, cov_matrix,%
  target,%
  skip,let,factor,observe,sample,return},
  morecomment={[s]{/*}{*/}},
}
\lstnewenvironment{lststan}%
{\lstset{language=Stan}}
{}
\lstnewenvironment{lststantable}%
{\lstset{language=Stan,basicstyle=\small\ttfamily,aboveskip=0em,belowskip=0em,xleftmargin=0em}}
{}
\lstnewenvironment{lststantablesmall}%
{\lstset{language=Stan,basicstyle=\fontsize{7}{8}\ttfamily,aboveskip=0em,belowskip=0em,xleftmargin=0em}}
{}
\def\stan{\lstinline[language=Stan, basicstyle=\normalsize\ttfamily]}

\begin{document}

\title{Automatic Guide Generation for Stan via NumPyro}

\author{Guillaume Baudart}
\affiliation{%
  \institution{Inria Paris, École normale supérieure -- PSL University}
}

\author{Louis Mandel}
\affiliation{%
  \institution{MIT-IBM Watson AI Lab, IBM Research}
}

\begin{abstract}
Stan is a very popular probabilistic language with a state-of-the-art HMC sampler but it only offers a limited choice of algorithms for black-box variational inference.
In this paper, we show that using our recently proposed compiler from Stan to Pyro, Stan users can easily try the set of algorithms implemented in Pyro for black-box variational inference.
We evaluate our approach on PosteriorDB, a database of Stan models with corresponding data and reference posterior samples.
Results show that the eight algorithms available in Pyro offer a range of possible compromises between complexity and accuracy.
This paper illustrates that compiling Stan to another probabilistic language can be used to leverage new features for Stan users, and give access to a large set of examples for language developers who implement these new features.
\end{abstract}

\maketitle

\section{Motivation}

The Stan probabilistic language~\cite{carpenter2017stan} enjoys broad adoption by the statistics and social sciences communities~\cite{gelman2006data, gelman2013bayesian,carlin2008bayesian}.
A Stan program defines a function from data and parameters to the value of a special variable \stan+target+ that represents the log-density of the model.
Given the observed data, the posterior distribution of the parameters can then be inferred using specialized inference algorithms, e.g., NUTS~\cite{homan2014nuts} (No U-Turn Sampler), an optimized Hamiltonian Monte Carlo (HMC) algorithm that is the preferred inference method for Stan.

Pyro~\cite{bingham_et_al_2019} and its JAX-based counterpart NumPyro~\cite{phan2019numpyro} on the other hand are generative probabilistic languages.
They describe \emph{generative models}, i.e., stochastic procedures that simulate the data generation process.
Generative PPLs are increasingly used in machine-learning research and are rapidly incorporating new ideas, such as Stochastic Gradient Variational Inference~(VI), or Bayesian neural networks in what is now called Deep Probabilistic Programming~\cite{bingham_et_al_2019, baudart_hirzel_mandel_2018, tran_et_al_2017}.

\bigskip
\emph{Variational Inference}
tries to find the member~$q_{\theta^*}(z)$ of a family~$\mathcal{Q} = \big\{q_{\theta}(z)\big\}_{\theta \in \Theta}$ of simpler distributions that is the closest to the true posterior $p(z \por \mathbf{x})$~\cite{blei_kucukelbir_mcauliffe_2017}.
Members of the family $\mathcal{Q}$ are characterized by the values of the \emph{variational parameters}~$\theta$.
The fitness of a candidate is measured using the Kullback-Leibler~(KL) divergence from the true posterior, which VI aims to minimize:
\begin{small}
$$
q_{\theta^*} (z)= \argmin_{\theta \in \Theta} \mbox{KL}\Big(q_{\theta}(z) \Par p(z \por \mathbf{x})\Big).
$$
\end{small}

Pyro natively supports variational inference and lets users define the
family $\mathcal{Q}$ (the \emph{variational guide}) alongside the model.
However, manually defining a correct and efficient guide is complex and error prone~\cite{Wonyeol_et_al_2020}.
Stan, on the other hand, offers ADVI~\cite{advi17jmlr}, an implementation of black-box VI where guides are automatically synthesized from the model using a mean-field or a full-rank approximation.
But, ADVI is only efficient for a subclass of models.

As an intermediate solution, Pyro developers recently introduced a zoo of \emph{autoguides}: variational guides that are automatically synthesized from the model using different heuristics~\cite{web2019pyrovi}.
Users can now try a range of synthesized guides on a given model before attempting to craft their own.

\bigskip

We recently proposed new backends for the Stanc3 Compiler to Pyro and NumPyro\footnote{\url{https://github.com/deepppl/stanc3}} and showed how to extend Stan with support for explicit variational guides~\cite{deepstan-short}.
In this paper, we show that our compiler and runtime can be used to test NumPyro autoguides on Stan models, and evaluate our approach on PosteriorDB a database of Stan models with corresponding data, and reference posterior samples~\cite{posteriordb}.

\section{Example}

\begin{figure}
\begin{minipage}[t]{0.45\linewidth}
\vspace{-3.35em}
\begin{lststantable}
parameters {
  real cluster; real theta; }
model {
  real mu;
  cluster ~ normal(0, 1);
  if (cluster > 0) mu = 20;
  else mu = 0;
  theta ~ normal(mu, 1); }
\end{lststantable}
\end{minipage}
\begin{minipage}{0.5\linewidth}
  \includegraphics[scale=0.7]{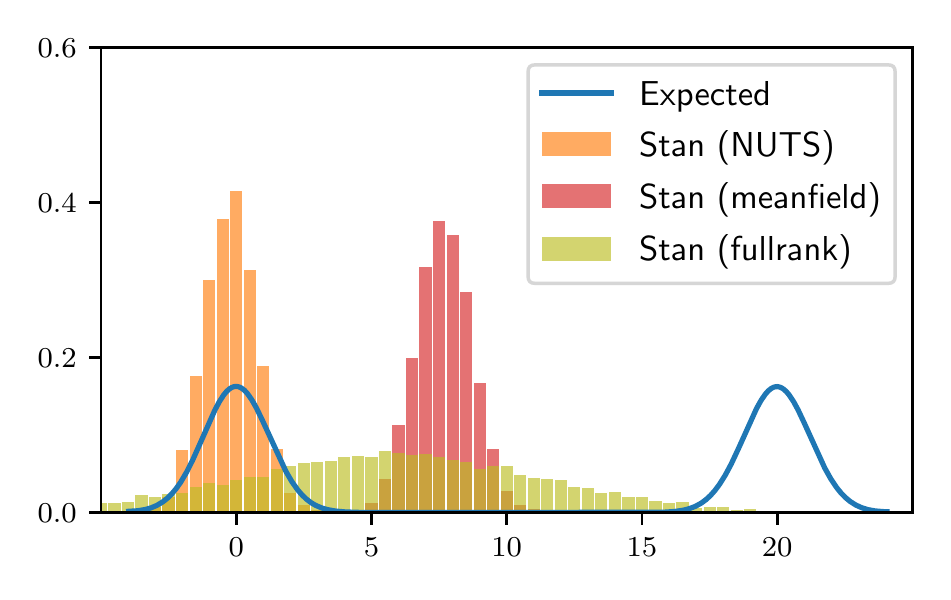}
  \vspace{-3em}
\end{minipage}
\vspace*{-0.4em}
\caption{\label{fig:multimodal}
Multimodal example in Stan, and graph of expected posterior distribution and histograms of the inferred posterior distributions using Stan NUTS and Stan ADVI with the mean-field and full-rank algorithms.}
\vspace*{-0.25em}
\end{figure}

The \emph{multimodal} example shown in \Cref{fig:multimodal} is a mixture of two Gaussian distributions with different means but identical variances.
This example is particularly challenging for NUTS.
Using multiple chains, NUTS finds the two modes, but the chains do not mix and the relative densities are incorrect.
This is a known limitation of HMC.\footnote{\url{https://mc-stan.org/users/documentation/case-studies/identifying_mixture_models.html}}
This model is also challenging for Stan's ADVI since the synthesized guide cannot approximate multi-modal distribution as illustrated in the histograms of \Cref{fig:multimodal} (by default, the runtime detects that ADVI does not converge and throws an exception). 

\medskip

Using our compiler from Stan to NumPyro we can try Pyro autoguides on the same example.
The following code illustrates our runtime.

\begin{lstpythontable}
from stannumpyro import NumPyroModel
import numpyro.infer.autoguide as autoguide
from numpyro.infer import Trace_ELBO
from numpyro.optim import Adam
from jax.random import PRNGKey

numpyro_model = NumPyroModel("multimodal.stan")
guide = autoguide.AutoBNAFNormal(numpyro_model.get_model())
svi = numpyro_model.svi(Adam(step_size=0.0005), Trace_ELBO(), guide)
svi.run(PRNGKey(0), {}, num_steps=10000, num_samples=10000)
\end{lstpythontable}

\noindent
The file \python{multimodal.stan} contains the Stan program of \Cref{fig:multimodal}.
From this file, we create a \python{NumPyroModel} object which compiles the model and loads the NumPyro code (line~7).
We then synthesize a guide for the model using the \python{AutoBNAFNormal} heuristic (line~8).
Following NumPyro's API, to launch the inference, we first create a SVI object from an optimizer (\python{Adam}), a loss function (\python{Trace_ELBO}), and the guide~(line~9).
Given a random seed \python{PRNGKEY(0)}, and the input data (here an empty dictionary since the model has no \stan{data} block), the \python{run} method first computes \python{num_steps} optimization steps and then draws \python{num_samples} samples from the posterior distribution~(line~10).

NumPyro offers eight different heuristics to synthesize a guide from the model:\footnote{\url{http://num.pyro.ai/en/stable/autoguide.html}}
\begin{itemize}
\item \python{AutoBNAFNormal} (\emph{Block Neural Autoregressive Flow}~\cite{bnaf}), 
\item \python{AutoDiagonalNormal} (Stan's mean-field), 
\item \python{AutoMultivariateNormal} (Stan's full-rank), 
\item \python{AutoIAFNormal} (\emph{Inverse Autoregressive Flow}~\cite{iaf}),
\item \python{AutoLaplaceApproximation} (quadratic approximation), 
\item \python{AutoLowRankMultivariateNormal}~\cite{lrvi},
\item \python{AutoNormal} (similar to \python{AutoDiagonalNormal}), and 
\item \python{AutoDelta} (MAP estimates).
\end{itemize}

\begin{figure}
\includegraphics[scale=0.75]{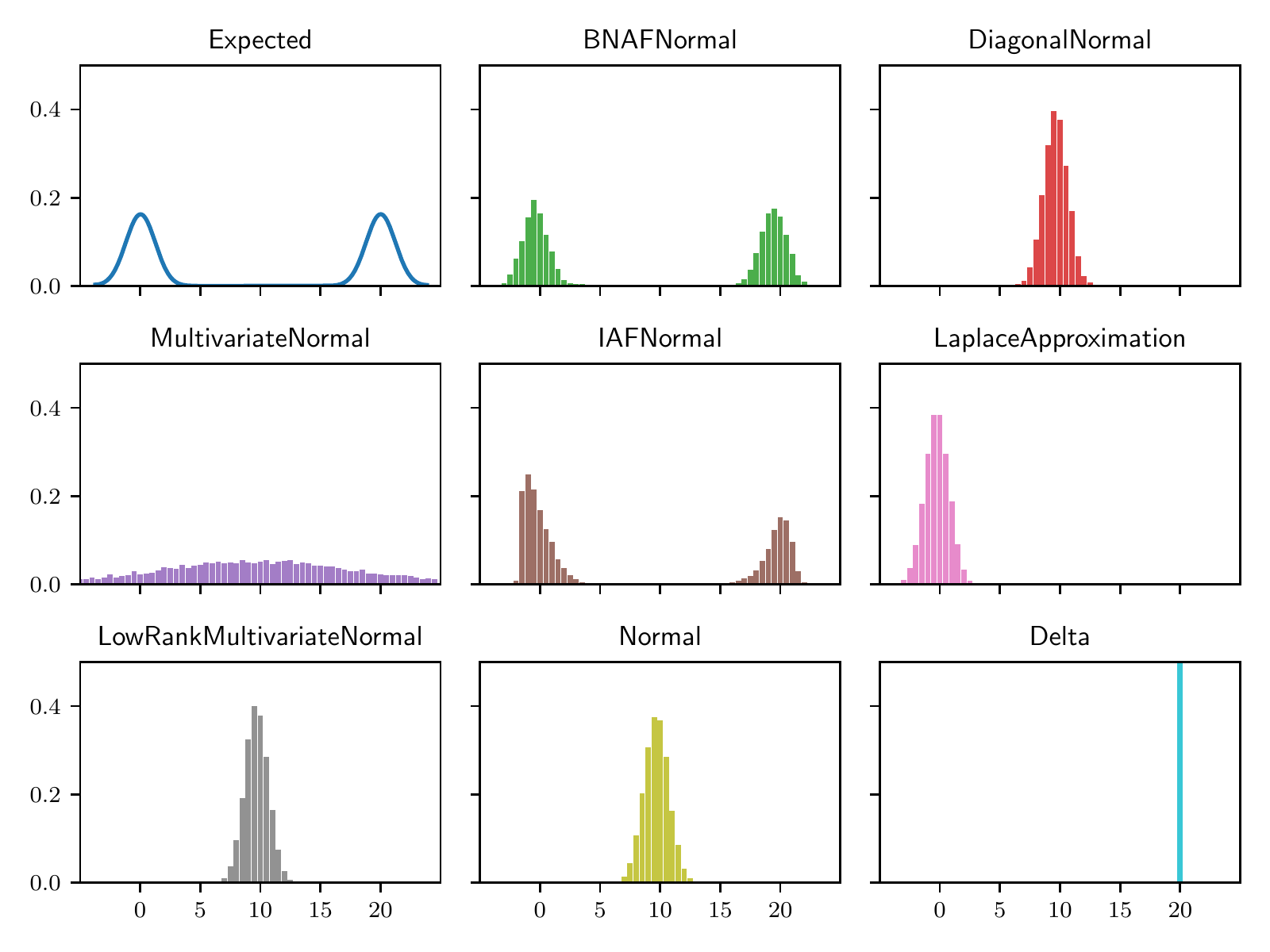}
\caption{Inference results using NumPyro autoguides on the model of \Cref{fig:multimodal} after 100,000 inference steps with 10,000 samples.}
\label{fig:autoguides_multimodal}
\end{figure}

\noindent
\Cref{fig:autoguides_multimodal} clearly shows that these heuristics yield different results on the example of \Cref{fig:multimodal}.
Only two of them successfully identify the two modes.

\section{Evaluation}

\begin{table}[]
\caption{Relative errors w.r.t. the reference samples of PosteriorDB. 
We report the maximum relative error across parameters in green if it is below~$0.3$, a \emark\ indicates a runtime error.}
\vspace{-1.5em}
\label{tab:eval_accuracy}
\begin{scriptsize}
\newcolumntype{R}[2]{%
    >{\adjustbox{angle=#1,lap=\width-(#2)}\bgroup}%
    l%
    <{\egroup}%
}
\newcommand*\rot{\multicolumn{1}{R{60}{1em}}}

\begin{sffamily}
\begin{tabular}{lrrrrrrrrrr@{}}
    \textsc{Dataset} - \textsc{Model} & \rot{BNAFNormal} & \rot{Delta} & \rot{DiagonalNormal} & \rot{IAFNormal} & \rot{LaplaceApproximation} & \rot{LowRankMultivariateNormal} & \rot{MultivariateNormal} & \rot{Normal}  & \rot{Stan MeanField} & \rot{Stan FullRank}\\
\cmidrule[\heavyrulewidth](lr){1-9}
\cmidrule[\heavyrulewidth](ll){10-11}
arK-arK                                     &            \Green{0.26} &       \Red{0.44} &                \Green{0.09} &           \Green{0.22} &                      \Red{0.38} &                           \Green{0.06} &                    \Green{0.13} &        \Green{0.17} &         \emark &      \Red{0.33} \\
arma-arma11                                 &              \emark &       \Green{0.26} &                  \emark &             \emark &                      \Green{0.16} &                             \emark &                      \emark &          \emark &         \emark &        \emark \\
sblri-blr                                   &            \Red{1.96} &       \Red{0.50} &                \Red{0.70} &           \Red{2.46} &                      \Red{0.38} &                           \Red{0.64} &                    \Red{0.76} &        \Red{0.64} &         \emark &        \emark \\
sblrc-blr                                   &            \Red{1.05} &       \Red{0.51} &                \Green{0.18} &           \Red{1.52} &                      \Red{0.40} &                           \Red{0.55} &                    \Red{1.58} &        \Green{0.27} &         \emark &        \emark \\
dogs-dogs                                   &            \Green{0.10} &       \Green{0.12} &                \Green{0.13} &           \Green{0.16} &                      \Green{0.11} &                           \Green{0.05} &                    \Green{0.08} &        \Green{0.13} &       \Red{0.38} &      \Red{0.80} \\
dogs-dogs\_log                               &              \emark &       \Green{0.12} &                \Green{0.07} &             \emark &                      \Green{0.12} &                           \Green{0.03} &                    \Green{0.05} &        \Green{0.07} &       \Red{0.93} &        \emark \\
earnings-earn\_height                        &            \Green{0.30} &      \Red{24.65} &               \Red{23.88} &             \emark &                     \Red{24.26} &                          \Red{24.21} &                   \Red{24.08} &       \Red{24.55} &         \emark &     \Red{18.71} \\
eight\_schools-eight\_schools\_centered        &            \Green{0.13} &       \Red{1.13} &                \Red{0.65} &           \Green{0.07} &                      \Red{1.13} &                           \Red{0.52} &                    \Red{0.60} &        \Red{0.62} &       \Red{1.17} &      \Red{1.39} \\
eight\_schools-eight\_schools\_noncentered     &            \Green{0.07} &       \Red{1.13} &                \Green{0.22} &           \Green{0.07} &                     \Red{13.03} &                           \Green{0.17} &                    \Green{0.18} &        \Green{0.26} &       \Green{0.22} &      \Green{0.21} \\
garch-garch11                               &            \Red{2.68} &       \Red{1.15} &                \Red{2.68} &           \Red{2.71} &                      \Red{0.74} &                           \Red{1.80} &                    \Red{1.68} &        \Red{2.69} &       \Red{0.81} &      \Red{0.34} \\
bball\_drive\_event\_0-hmm\_drive\_0             &            \Green{0.28} &     \Red{290.68} &                \Green{0.29} &             \emark &                      \Red{0.42} &                           \Red{0.43} &                    \Green{0.27} &        \Green{0.10} &       \Red{0.48} &     \Red{14.31} \\
bball\_drive\_event\_1-hmm\_drive\_1             &            \Red{0.53} &       \Red{0.59} &                \Red{0.52} &           \Red{0.77} &                    \Red{768.80} &                           \Red{0.52} &                    \Red{0.63} &        \Red{0.48} &       \Red{9.68} &      \Red{4.59} \\
hmm\_example-hmm\_example                     &            \Green{0.16} &       \Red{0.37} &                \Red{0.46} &             \emark &                      \Green{0.25} &                           \Green{0.05} &                    \Green{0.30} &        \Red{0.36} &       \Red{1.73} &      \Red{0.55} \\
kidiq-kidscore\_interaction                  &            \Red{0.62} &       \Red{0.63} &                \Red{2.47} &           \Red{0.82} &                      \Red{0.49} &                           \Red{2.46} &                    \Red{2.21} &        \Red{2.47} &       \Red{2.61} &        \emark \\
kidiq\_with\_mom\_work-kidscore\_interaction\_c  &            \Red{0.73} &      \Red{48.31} &               \Red{46.66} &           \Green{0.23} &                     \Red{46.42} &                          \Red{48.03} &                   \Red{49.68} &       \Red{46.71} &      \Red{44.70} &     \Red{56.46} \\
kidiq\_with\_mom\_work-kidscore\_interaction\_c2 &            \Red{0.67} &      \Red{35.30} &               \Red{34.10} &           \Red{0.33} &                     \Red{33.91} &                          \Red{34.35} &                   \Red{35.09} &       \Red{34.13} &         \emark &     \Red{93.98} \\
kidiq\_with\_mom\_work-kidscore\_interaction\_z  &            \Red{0.70} &      \Red{48.03} &               \Red{46.41} &           \Red{0.44} &                     \Red{46.15} &                          \Red{48.49} &                   \Red{49.75} &       \Red{46.45} &       \Red{2.87} &     \Red{48.65} \\
kidiq\_with\_mom\_work-kidscore\_mom\_work       &            \Red{0.65} &      \Red{16.47} &               \Red{15.94} &           \Green{0.17} &                     \Red{15.79} &                          \Red{16.05} &                   \Red{16.04} &       \Red{15.95} &       \Red{1.94} &      \Red{7.23} \\
kidiq-kidscore\_momhs                        &            \Green{0.19} &      \Red{16.61} &               \Red{15.84} &             \emark &                     \Red{15.66} &                          \Red{15.96} &                   \Red{16.02} &       \Red{15.83} &       \Red{6.76} &      \Red{2.18} \\
kidiq-kidscore\_momhsiq                      &            \Green{0.27} &       \Green{0.17} &                \Red{2.44} &           \Green{0.25} &                      \Green{0.12} &                           \Red{2.28} &                    \Red{1.82} &        \Red{2.44} &         \emark &      \Red{4.07} \\
kidiq-kidscore\_momiq                        &            \Green{0.10} &       \Green{0.15} &                \Red{2.20} &             \emark &                      \Green{0.12} &                           \Red{2.13} &                    \Red{1.75} &        \Red{2.21} &       \Red{3.28} &      \Red{3.65} \\
kilpisjarvi\_mod-kilpisjarvi                 &            \Red{2.12} &       \Red{1.99} &                \Red{1.97} &             \emark &                      \Red{1.95} &                           \Red{1.98} &                    \Red{7.00} &        \Red{1.98} &         \emark &        \emark \\
earnings-log10earn\_height                   &            \Green{0.12} &       \Green{0.09} &                \Green{0.17} &             \emark &                      \Green{0.08} &                           \Green{0.15} &                    \Red{0.56} &        \Green{0.19} &         \emark &        \emark \\
earnings-logearn\_height                     &            \Green{0.13} &       \Green{0.09} &                \Red{1.08} &             \emark &                      \Green{0.08} &                           \Red{1.11} &                    \Red{1.21} &        \Red{1.08} &         \emark &        \emark \\
earnings-logearn\_height\_male                &            \Green{0.25} &       \Green{0.11} &                \Red{4.22} &           \Green{0.16} &                      \Green{0.08} &                           \Red{4.06} &                    \Red{3.74} &        \Red{4.21} &       \Red{1.33} &     \Red{10.73} \\
earnings-logearn\_interaction                &            \Red{0.56} &       \Green{0.12} &                \Red{5.52} &           \Red{0.81} &                      \Green{0.11} &                           \Red{5.55} &                    \Red{5.99} &        \Red{5.51} &         \emark &        \emark \\
earnings-logearn\_interaction\_z              &            \Red{0.67} &       \Green{0.12} &                \Green{0.27} &           \Red{0.42} &                      \Green{0.10} &                           \Green{0.10} &                    \Green{0.27} &        \Green{0.15} &       \Red{1.90} &      \Red{1.28} \\
earnings-logearn\_logheight\_male             &            \Red{0.45} &       \Green{0.12} &                \Red{1.22} &           \Red{0.54} &                      \Green{0.09} &                           \Red{1.21} &                    \Red{1.17} &        \Red{1.22} &       \Red{0.66} &        \emark \\
mesquite-logmesquite                        &            \Green{0.15} &       \Green{0.93} &                \Green{0.03} &           \Green{0.22} &                      \Red{0.80} &                           \Green{0.05} &                    \Green{0.06} &        \Green{0.05} &         \emark &      \Green{0.11} \\
mesquite-logmesquite\_logva                  &            \Red{0.34} &       \Red{0.65} &                \Green{0.06} &           \Red{0.37} &                      \Red{0.52} &                           \Green{0.05} &                    \Green{0.05} &        \Green{0.05} &       \Green{0.14} &      \Green{0.05} \\
mesquite-logmesquite\_logvas                 &            \Green{0.14} &       \Red{0.93} &                \Green{0.13} &           \Green{0.27} &                      \Red{0.80} &                           \Green{0.11} &                    \Green{0.06} &        \Green{0.13} &         \emark &      \Green{0.09} \\
mesquite-logmesquite\_logvash                &            \Green{0.22} &       \Green{0.84} &                \Green{0.08} &           \Green{0.26} &                      \Red{0.72} &                           \Green{0.06} &                    \Green{0.08} &        \Green{0.05} &         \emark &        \emark \\
mesquite-logmesquite\_logvolume              &            \Red{0.32} &       \Red{0.45} &                \Green{0.13} &             \emark &                      \Red{0.33} &                           \Green{0.11} &                    \Green{0.13} &        \Green{0.04} &       \Green{0.22} &      \Green{0.08} \\
low\_dim\_gauss\_mix-low\_dim\_gauss\_mix         &            \Red{0.68} &       \Green{0.13} &                \Green{0.22} &           \Red{0.75} &                      \Green{0.08} &                           \Green{0.21} &                    \Green{0.23} &        \Green{0.27} &      \Red{63.08} &     \Red{62.71} \\
mesquite-mesquite                           &            \Green{0.16} &       \Red{8.12} &               \Red{11.02} &           \Red{6.39} &                      \Red{7.89} &                          \Red{11.80} &                   \Red{11.25} &       \Red{10.94} &      \Red{23.32} &     \Red{17.18} \\
nes1988-nes                                 &            \Green{0.18} &       \Green{0.27} &                \Green{0.06} &           \Red{0.33} &                      \Green{0.24} &                           \Red{0.33} &                    \Green{0.06} &        \Green{0.11} &       \Red{0.88} &      \Red{2.09} \\
nes1972-nes                                 &            \Green{0.19} &       \Green{0.21} &                \Green{0.05} &           \Red{0.33} &                      \Green{0.18} &                           \Red{0.38} &                    \Green{0.05} &        \Green{0.15} &       \Green{0.27} &      \Red{1.30} \\
nes1984-nes                                 &            \Green{0.16} &       \Green{0.23} &                \Green{0.04} &           \Green{0.30} &                      \Green{0.20} &                           \Red{0.36} &                    \Green{0.04} &        \Green{0.14} &       \Red{1.25} &      \Red{1.31} \\
nes1980-nes                                 &            \Green{0.14} &       \Green{0.29} &                \Green{0.03} &           \Green{0.19} &                      \Green{0.25} &                           \Red{0.31} &                    \Green{0.03} &        \Green{0.13} &       \Red{1.05} &      \Red{0.74} \\
nes2000-nes                                 &            \Green{0.12} &       \Red{0.37} &                \Green{0.04} &           \Green{0.19} &                      \Red{0.32} &                           \Green{0.25} &                    \Green{0.03} &        \Green{0.12} &       \Red{1.01} &      \Red{1.28} \\
nes1976-nes                                 &            \Green{0.19} &       \Green{0.24} &                \Green{0.05} &           \Red{0.32} &                      \Green{0.21} &                           \Red{0.35} &                    \Green{0.06} &        \Green{0.13} &       \Red{2.61} &      \Red{1.89} \\
nes1992-nes                                 &            \Green{0.18} &       \Green{0.23} &                \Green{0.05} &           \Green{0.27} &                      \Green{0.21} &                           \Red{0.35} &                    \Green{0.05} &        \Green{0.13} &       \Red{2.19} &      \Red{0.85} \\
nes1996-nes                                 &            \Green{0.17} &       \Green{0.25} &                \Green{0.04} &           \Green{0.25} &                      \Green{0.21} &                           \Red{0.34} &                    \Green{0.05} &        \Green{0.14} &       \Red{1.29} &      \Red{2.54} \\
\cmidrule(lr){1-9}
\cmidrule(ll){10-11}
\textsc{Average}    &                   0.46 &             11.72 &                       5.30 &                  0.71 &                            22.89 &                                  5.43 &                           5.59 &               5.32 &              6.16 &            11.30 \\
\textsc{Successes} &                     25 &                21 &                         22 &                    16 &                               20 &                                    14 &                             21 &                 22 &                 4 &                5 \\
\textsc{Mismatches} &                     16 &                22 &                         20 &                    16 &                               23 &                                    28 &                             21 &                 20 &                25 &               27 \\
\textsc{Errors}   &                      2 &                0 &                          1 &                    11 &                                 0 &                                     1 &                              1 &                  1 &                14 &               11 \\
\cmidrule[\heavyrulewidth](lr){1-9}
\cmidrule[\heavyrulewidth](ll){10-11}
\end{tabular}
\end{sffamily}
\end{scriptsize}
\end{table}

PosteriorDB~\cite{posteriordb} provides reference samples for $49$ pairs (model, dataset).
Due to missing functions in our implementation of the standard library (e.g., ODE solvers), we cannot test six models.
For each of the remaining $43$ models, we run 100,000 inference steps (using \python{Adam(step_size=0.0005)}) and generate 10,000 samples from the posterior distribution.

\pagebreak

To evaluate inference accuracy, for each parameter $x$ in the posterior distribution we compute the relative error w.r.t. to the reference samples.
For multidimensional parameters we compute the error for every component.
$$
  \mathit{err} = \frac{| \textrm{mean}(x_{\textrm{ref}}) - \textrm{mean}(x)|}{\textrm{stddev}(x_{\textrm{ref}})}
$$

The evaluation code is available at \url{https://github.com/deepppl/evaluation-autoguide}.
\Cref{tab:eval_accuracy} summarizes the results.
For each model, we report the maximal relative error across parameters.
Relative errors below $0.3$ --- the criteria used by regression tests for Stan\footnote{\url{https://github.com/stan-dev/performance-tests-cmdstan}} --- appear in green.
On this set of benchmarks, the autoguides relying on normalizing flows \python{AutoBNAFNormal} and \python{AutoIAFNormal} outperform other autoguides.
\python{AutoDelta} can be used with all models without raising runtime errors, but it only computes a MAP estimate.
\python{AutoLaplaceApproximation} also runs without error on all models but returns less accurate distributions based on a MAP estimations.
Runtime errors correspond to NaN values.
Overall, the autoguides demonstrate a range of possible compromises between complexity and accuracy.

For comparison, we also report the results of Stan ADVI with the mean-field and full-rank algorithms.
Compared to NumPyro, instead of running a fixed number of iterations, the Stan runtime relies on an adaptive optimization sequence and throws an exception when the algorithm fails to converge.

\section{Conclusion}

In this paper, we showed that by compiling Stan models to NumPyro, Stan users now have access to a large variety of automatically synthesized guides for variational inference.
As illustrated on 43 benchmarks from PosteriorDB, these new guides offer new compromises between complexity and accuracy compared to Stan ADVI full-rank and mean-field algorithms.\footnote{\url{https://discourse.mc-stan.org/t/intermediate-between-mean-field-and-full-rank-advi}}

This paper illustrates that compiling Stan to another probabilistic language can be used to leverage new features for Stan users, and give access to a large set of examples for language developers who implement these new features.

\begin{acks}
We would like to thank Eli Bingham, Fritz Obermeyer, and Du Phan for suggesting us this application of the Stan to NumPyro compiler.
\end{acks}

\bibliographystyle{acm-reference-format}
\bibliography{bibfile}

\end{document}